\documentclass{article}

\usepackage{arxiv}

\usepackage[utf8]{inputenc}         % allow utf-8 input
\usepackage[T1]{fontenc}            % use 8-bit T1 fonts
\usepackage[hidelinks]{hyperref}    % hyperlinks
\usepackage{url}                    % simple URL typesetting
\usepackage{booktabs}               % professional-quality tables
\usepackage{amsfonts}               % blackboard math symbols
\usepackage{nicefrac}               % compact symbols for 1/2, etc.
\usepackage{microtype}              % microtypography
\usepackage{lipsum}
\usepackage{mathtools}
\usepackage{enumitem}
\usepackage{tikz}
\usepackage{pgf-umlsd}
\usepackage{scalefnt}
\usepackage{float}

\title{How to Authenticate MQTT Sessions without Channel- and Broker Security}

\author{
  Reto E. Koenig\\
  Department of Computer Science\\
  Bern University of Applied Sciences\\
  Switzerland\\
  \texttt{reto.koenig@bfh.ch} \\
  \And
  Lukas Laederach \\
  Student \\
  Bern University of Applied Sciences\\
  Switzerland\\
  \texttt{laederach.lukas@gmail.com} \\
  \And
  Cédric von Allmen \\
  Student \\
  Bern University of Applied Sciences\\
  Switzerland\\
  \texttt{cedric.vonallmen@gmail.com} \\
}

\begin{document}
\maketitle

\begin{abstract}
  This paper describes a new but state of the art approach to provide authenticity in MQTT sessions using the means of Zero-Knowledge Proofs. This approach completely voids session hijacking for the MQTT protocol and provides authenticity without the need for any network security nor channel security nor broker based predefined Access Control List (ACL). The presented approach does not require the broker to keep any secrets for session handling. Moreover, it allows the clientID, which represents the identification for a session, to be publicly known. The presented approach allows completely anonymous but authentic sessions, hence the broker does not need any priory knowledge of the client party. As it is especially targeted for applications within the world of Internet of Things (IoT), the presented approach is designed to require only the minimum in extra power in terms of energy and space. The approach does not introduce any new concept, but simply combines a state of the art cryptographic Zero-Knowledge Proof of identity with the existing MQTT 5.0 specification. Thus, no protocol extension is required in order to provide the targeted security properties. The described approach is completely agnostic to the application layer at the client side and is only required during MQTT session establishment.
\end{abstract}

% keywords can be removed

\section{Introduction}
\label{theo:authn:introduction}
MQTT is designed as a robust, session-oriented protocol especially suitable for the world of IoT, where the clientID plays the central role for session management. The MQTT specification requires the clientID to be provided within the first data frame of the protocol during session establishment. The semantics of the clientID is to provide the unique way a session can be (re)established between a client and the broker, without any further information. So the clientID is required to be unique per broker over time, hence, no collision of clientIDs should ever happen. As the clients are not aware of each other, but usually provide their own clientID, it must be drawn from large set of possible clientIDs so the probability of a collision of clientIDs is negligible. The protocol specification defines a minimum clientID space of order $63^{23}$. Hence, allows the theoretical security parameter of $2^\lambda$ with an approximate $\lambda=137$. 

The specification of MQTT does not directly address the immanent possibility of active session hijacking. This attack vector allows an adversary to take over the session by simply (re-)establishing a connection using the clientID of the victim. Only indirectly, the MQTT specification tries to weaken that immanent attack vector by providing the possibility for channel security via Transport Layer Securit (TLS) or even worse, by securing a whole network via Virtual Private Network (VPN), and of course the heavy security requirement of the broker to be fully trusted in terms of secrecy of the clientID. However, the MQTT protocol specification gives room for customized authentication and even provides protocol intrinsic authentication solution via simple username password or via ACLs for ID-based session handling. These ID-based approaches, however, pose several disadvantages in the context of client management and security, as they require a separate and out-of-band on-staging-phase comprising the broker in order to manage clients by the identification scheme. Moreover, all these methods require the broker to keep secrets shared amongst different clients and the broker.

The presented approach provides a fully anonymous identification scheme based on the Schnorr Non-Interactive Zero-Knowledge Proof (Schnorr NIZKP) identification scheme \cite{rfc8235} without any of the drawbacks mentioned above. It makes use of well-known cryptographic properties and allows the security model of the broker to be lowered from "fully trusted" to "honest but curious... without any privacy constraints". Again, there is no new concept introduced, but well-known concepts are simply combined in order to provide the desired property.

The following sections will introduce the security and adversary model this approach is operating in. This way, a discussion about the targeted adversary and the trust and security assumptions is possible. Then the cryptographic and protocol prerequisites are provided in order to understand the implementation that follows. The paper concludes by providing some quantitative and qualitative analysis of the approach.

\section{Security model}
\label{theo:authn:securitymodel}
The main security parameter is denoted by $\lambda \in \mathbb{N}$. We write $a \leftarrow A(x)$ if $a$ is assigned to the output of algorithm $A$ with input $x$. An algorithm is efficient if it runs in probabilistic polynomial time (\emph{ppt}) in the length of its input. For the remainder of this paper, all algorithms are ppt if not explicitly mentioned otherwise. If $\emph{S}$ is a set, we write $a \leftarrow_R \emph{S}$ to denote that $a$ is chosen uniformly at random from $\emph{S}$. For a message $m = (m[1], m[2], \ldots , m[l])$, we call $m[i]$ a block, while $l \in \mathbb{N}$ denotes the number of blocks in a message $m$. For a list we require that we have unique, injective, and efficiently reversible encoding, which maps the list to ${0, 1}^*$.

The security model that covers the approach consists of three actors, namely the channel, the broker and the client. There are no security assumptions required for the channel except of eventual availability. The channel is allowed to be modelled as an unreliable broadcast channel. The memory of the broker can be modelled as public readable memory, thus no secrecy is required on the broker side concerning the management of the clientIDs. The client is required to be able to keep a secret which might be either completely intrinsic to the client e.g. provided by a Physical Unclonable Function (PUF), or injected via application level.

The security properties of the actors define the adversary model. The adversary which is assumed to be restricted to be in ppt only is allowed to have full knowledge about the channel and the memory of the broker at any time (\emph{full-take}). The adversary, however, is not able to extract the secret from the client.

\section{Prerequisites}
\label{theo:authn:prerequisits}
In the following section the prerequisites are described in order to understand the solution. As this solution proposes a cryptographic approach within the given protocol specification, both aspects are described here.

\subsection{Cryptographic prerequisites}
\label{theo:authn:cryptograhicprerequisits}
The cryptographic protocol used for this solution is the Schnorr identification scheme as described in the RFC-8235. In order to render this paper self-contained, a brief summary of the RFC-8235 is given here. Please refer to the full RFC document for further details \cite{rfc8235}.

The Schnorr NIZKP can be implemented over a finite field or an elliptic curve. The technical specification is basically the same, except that the underlying cyclic group is different. For simplicity, this document describes the approach within the finite field, whereas in the RFC-document, both versions are described.

Let $p$ and $q$ be two large primes with $q | p-1$.  Let $\mathbb{G}_q$ denote the subgroup of $\mathbb{Z}_p^*$ of prime order $q$, and $g$ be a generator for the subgroup.
\begin{description}
\item[Statement] $\textit{NIZKP}[(x): y = g^{x}]$ Alice knows the discrete logarithm $x$ of the value $y$ to the base $g$ (within $\mathbb{Z}_p^*$).
\item[Public input] $p,q,g,\mathbb{Z}^*_p, \mathbb{G}_q$, and the cryptographic hash function $H(x)$.
\item[Prover's (Alice) private input.]  $x \in \mathbb{Z}_q$ such that $y=g^x \mod{p}$.
\item[P $\xrightarrow{(t,s)}$ V] Alice chooses random $\omega \leftarrow_R \mathbb{Z}_q$, calculates $t=g^\omega$, calculates $c = H(g || t || y || \textit{public data})$,\\ calculates $s = \omega + x \cdot c \mod{q}$
\item[Verification.] Bob \emph{accepts} the proof only if $y \in \mathbb{Z}^*_p$ and $g^s = t \cdot y^c  \mod{p}$.
\end{description}

\subsection{Protocol prerequisites}
\label{theo:authn:protocolprerequisits}
The MQTT protocol \cite{MQTTVers5:online} describes the possibility for authentication exchange in section 3.15 and section 4.12. The authors of the MQTT protocol were aware of the problem of session hijacking and replay attacks. Especially in section 5.4.5 of the protocol specification, they give hints on how to mitigate the attack attempts. But all these mitigation approaches are not focusing on the problem at hand but augment the security assumptions of the channel or even of the network, by requiring the channel to be secured (\emph{TLS}) or even by asking for a secure network environment (\emph{VPN}).

As already advertised in the security model, the approach does not require such heavy security assumptions. It is based on a challenge response mechanism already foreseen by the specification, very much like the described Salted Challenge Response Authentication Mechanism (\emph{SCRAM}) in the non-normative part of the protocol description. Thus the approach perfectly fits within the given protocol boundaries.

\section{Protocol description}
\label{theo:authn:protocoldescription}
The main idea of this protocol is to choose the $\textit{clientID}$ as a group element $y \in \mathbb{G}_q$. In the notation for the multiplicative group, the client chooses $y=g^x$ with $x \leftarrow_R \mathbb{Z}_q$. In order to void replay attacks, the broker provides  the client with a nonce $n$ to be used as \emph{public data} for the Non-interactive Zero-Knowledge Proof. Only if the client is able to provide the proof $(t,s)$ within the same network connection as the provided challenge, the broker proceeds with  a CONNACK containing the reason code \textit{0x00 (Success)}, in all other cases it disconnects with $\textit{DISCONNECT}$ and some reason code e.g. \textit{0x83 (Implementation specific error)}. This way, the client provides an anonymous identification each time it establishes a fresh connection.

\section{Implementation}
\label{theo:authn:implementation}
A first approach would be a general purpose Schnorr NIZKP system, but the generality comes with a price, in the form of substantial processing time and large public parameters required to construct a proof \cite{arxiv:a-nizkp-ctf-platform:online}. 

The approach proposed in this paper consists in choosing a key pair $(sk, pk)$ which verifies whether $sk$ is the private key corresponding to the public key $pk$ in a digital signature scheme. This choice allows us to reduce our proof of knowledge problem to that of digitally signing messages, whose implementation is simpler and more efficient than any known general purpose Schnorr NIZKP system \cite{arxiv:a-nizkp-ctf-platform:online}. 

Let the following be the primitives of a secure digital signature scheme:

\begin{equation}
    \label{theo:authn:implementation:eq:sign}
    Sign(sk, m)= s||m
\end{equation}
\begin{equation}
    \label{theo:authn:implementation:eq:verify}
    Verify(pk, s||m)=
    \begin{cases}
        m, & \text{if s is valid}\ \\
        \bot, & \text{otherwise}
    \end{cases}
\end{equation}

\autoref{theo:authn:implementation:eq:sign} signs the message $m$ using the private key $sk$, and prepends the signature $s$ to the message $m$. \autoref{theo:authn:implementation:eq:verify} verifies whether $s$ is a valid signature for $m$ produced by the private key $sk$ corresponding to the public key $pk$ then outputs the original message $m$ if the signature is valid, or $\bot$ if it is invalid \cite{arxiv:a-nizkp-ctf-platform:online}. 

It is assumed that the given digital signature scheme satisfies completeness, validity and zero-knowledge to be useful in this context.

The implementation requires some addition to the client and broker software, that provide the MQTT ability.

\begin{figure}[H]
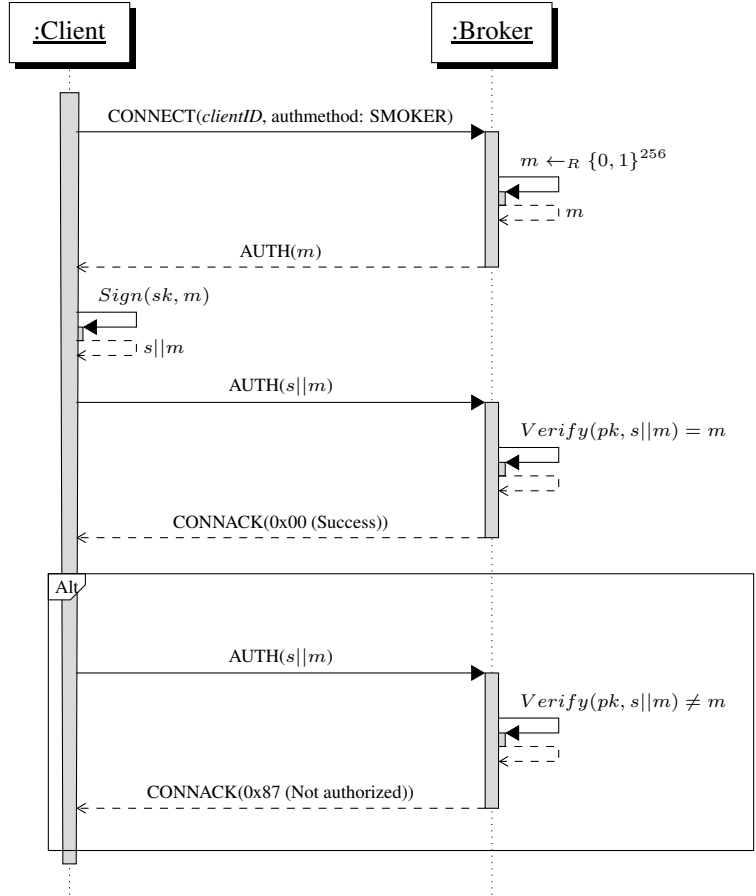

  \centering
  \begin{sequencediagram}
  \scalefont{1}
    \newthread{client}{:Client}{client}
    \newinst[4]{broker}{:Broker}{broker}

    \scalefont{0.7}

    %connect & get nonce from server
    \begin{call}{client}{CONNECT($\textit{clientID}$, authmethod: SMOKER)}{broker}{AUTH($m$)}

      % genereate nonce
      \begin{callself}{broker}{$m \leftarrow_R \{0,1\}^{256}$}{$m$}
      \end{callself}

    \end{call}

    % create signature
    \begin{callself}{client}{$Sign(sk, m)$}{$s||m$}
    \end{callself}

    % send prood to broker
    \begin{call}{client}{AUTH($s||m$)}{broker}{CONNACK(0x00 (Success))}
      
      % verify signature
      \begin{callself}{broker}{$Verify(pk, s||m) = m$}{}
      \end{callself}

    \end{call}
    \begin{sdblock}{Alt}{}
      \begin{call}{client}{AUTH($s||m$)}{broker}{CONNACK(0x87 (Not authorized))}
      
        % verify signature
        \begin{callself}{broker}{$Verify(pk, s||m) \ne m$}{}
        \end{callself}
  
      \end{call}
    \end{sdblock}
  \end{sequencediagram}
  \caption{Successful MQTT session establishment using the SMOKER authentication method.}
  \label{theo:authn:fig:succ:sess:estab}
\end{figure}

The following \autoref{theo:authn:fig:succ:sess:estab} visualizes a the enhanced authentication process. For further examples within this paper the authentication method, which is needed to trigger enhanced authentication on the broker, is named \textit{SMOKER}. The process uses the signature primitives mentioned in \autoref{theo:authn:implementation:eq:sign} and \ref{theo:authn:implementation:eq:verify} and the key pair $(sk, \textit{clientID} \leftarrow_{mqtt} pk)$. As $pk$ must serve as clientID, it must get mapped into the set of MQTT-$\textit{clientID}$s. ClientIDs are further analyzed in \autoref{theo:authn:implementation:analyses}. 

% What is the client and broker doing here in terms of frames (Sequence-Diagram)

\paragraph{Client}
Prior to the connection the client calculates its $\textit{clientID}$ by deriving a value $pk$ from its secret key $sk$ (can be the result of a puf or an out-of-band injected value). This is then mapped into the to the allowed set of MQTT-$\textit{clientID}$s  $\textit{clientID} \leftarrow_{mqtt} pk$. After the client established a connection to the broker, the client starts a MQTT session by sending the connect frame containing the authentication flag in expectancy of a broker authentication response, where a nonce $m \leftarrow_R \{0,1\}^{256}$ has to be provided. The client signs the received nonce using the secure signature scheme. 

The client then sends the signed nonce $s||m$ as an authentication response within the same connection to the broker. If the client needs to re-establish the session, the very same procedure is executed as if the client would establish the session for the very first time.

\paragraph{Broker}
If a client is connecting to the broker, the provided $\textit{clientID}$ is only accepted, once the authentication succeeds within the same connection. The broker generates a random nonce $m \leftarrow_R \{0,1\}^{256}$ as a response to the first authentication frame within a new connection. The response delivered from the client consists of the signed nonce. The broker verifies the signature and thus extracts $m'$ from $s||m$ which must be equal to the originally transmitted $m$. Only if this verification is successful, the $\textit{clientID}$ is accepted and the broker behaves in its usual way.

\subsection{Connecting with an unknown authentication method}
As shown in \autoref{theo:authn:fig:sess:estab:wrong:auth}, the given authentication method, requested by the client, must be supported by the broker -- otherwise a CONNACK with the reason code 0x8C (Bad authentication method) \cite[Table 3‑1]{MQTTVers5:online} must be returned and the connection is closed by the broker. An empty authentication method is not further handled by the extended authentication mechanism and is delegated to the brokers default implementation.

% Sequencediagram
\newcommand{\newspacedthread}[3][0.2]{
  \newinst[#1]{#2}{#3}
  \stepcounter{threadnum}
  \node[below of=inst\theinstnum,node distance=0.8cm] (thread\thethreadnum) {};
  \tikzstyle{threadcolor\thethreadnum}=[fill=gray!30]
  \tikzstyle{instcolor#2}=[fill=gray!30]
}

\begin{figure}[H]
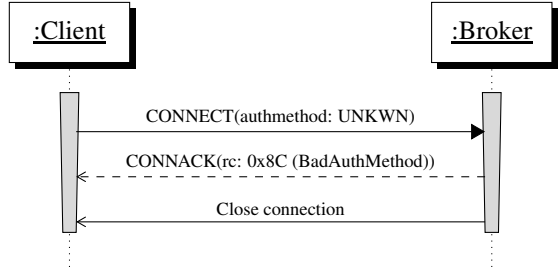

  \centering
  \begin{sequencediagram}
    \newthread{client}{:Client}{client}
    \newspacedthread[4]{broker}{:Broker}{broker}

    \scalefont{0.7}

    %connect & get nonce from server
    \begin{call}{client}{CONNECT(authmethod: UNKWN)}{broker}{CONNACK(rc: 0x8C (BadAuthMethod))}
    \end{call}

    % if rejected -> broker closes connection
    \mess{broker}{Close connection}{client}

  \end{sequencediagram}
  \caption{MQTT connect with an unknown authentication method.}
  \label{theo:authn:fig:sess:estab:wrong:auth}
\end{figure}

\subsection{ClientID stealing}
\label{theo:authn:implementation:id:stealing}
Common broker implementations are disconnecting an active client session if another client is connecting with the same clientID. The specification is very clear  \cite[section 3.1.3.1]{MQTTVers5:online}:

\begin{quotation}
    \textit{
        ``The broker must not accept connections with the same clientID''
    }
\end{quotation}

As the clientID does not need to be treated as a secret anymore, the broker must ensure that an unauthenticated client is not able to steal a clientID of a proper authenticated client. In this case the broker must reject the connection by sending a CONNACK packet with a \textit{0x85 (Client Identifier not valid)} reason code. Conversely, a connecting client that successfully authenticates, must be preferred over any other unauthenticated client with the same clientID -- the unauthenticated client must get disconnected immediately.

\section{Analyses}
\label{theo:authn:implementation:analyses}
First a fair warning: Due to the 'short' maximum length of the $\textit{clientID}$ covered by the MQTT specification in relation to cryptographic constraints, it is important to implement the proof or signature using the mathematics of elliptic curves. This way, the security parameter $\lambda$ provided by the MQTT specification provides a maximum $\lambda=70$. According to cryptographic sources \cite{Gir08}, this is already a very low security. But using finite fields would lower $\lambda$ by at least another factor of $20$ and would not sustain an attack of any computational moderate potent adversary. Therefore, as an advice, the client and the broker must be enabled to accept a clientID set of order $ \geq 2^{256}$. Using the alphabet supported by the MQTT specification this results in a minimum of $43$ characters for the clientID in order to get a minimum $\lambda=128$.

\subsection{Energy and space requirements}
\label{theo:authn:energyspacerequirements}
In terms of computing power, the client must be able to execute two operations on an elliptic curve and one cryptographic hash per connection. The expensive calculations on the elliptic curve can even be lowered down to a single operation if the $\textit{clientID}$ derived from the according secret can be stored in the system. The computing power on the broker side must allow to calculate a minimum of two operations on the elliptic curve and one cryptographic hash per connection. Furthermore, the broker is required to provide a high entropy challenge nonce $m$ per connection. Either, the broker gets true randomization, which is a very difficult task, or the broker uses an initialization vector (secret iv) seed to initialize a pseudo random generator function. It can keep seeding the pseudo random generation function by the answers it receives from the client connections. This way, the entropy of the client secret can be used to influence the available entropy on the broker, where natural sources of entropy are sparse by nature.

\subsection{Passive adversary}
The provided implementation allows the broker to reside in a very relaxed security model. No secret is required anymore concerning the $\textit{clientID}$. The provided implementation further does not require any special treatment of the channel used for communication, as no secret is sent over the channel at all. As long as the underlying network protocol provides authenticity in terms of source and destination during connection, the MQTT session cannot be hijacked at all. The only remaining party within a strict security model is the client. It must be able to keep the secret safe from being extracted.

\subsection{Active adversary}
As long as the adversary is not in possession of the secret $x$ kept by the client, it cannot provide a valid proof of knowledge to the broker. However, if the adversary model is relaxed, it can start a Man In The Middle (MITM) attack. This allows the adversary to impersonate the client without the client nor the broker to gain any knowledge of that. Thus, the client is required to be sure about the public identity of the desired broker. This brings us back to TLS (and all its pros and cons) with server certificate, in order to provide server-identity 'only' in order to prevent the man-in-the-middle attack. So, one might argue, that the title of this paper might be miss leading, but it must be stressed out, that TLS is not required to prevent clientID stealing, but for broker identity in order to prevent the active adversaries MITM attack.

\bibliographystyle{IEEEtran}  
\bibliography{bibliography}

\end{document}